\documentclass[preprint,prb]{revtex4}
\usepackage{graphicx}
\begin{document}

\title{NMR study on the stability of the magnetic ground state in MnCr${}_2$O${}_4$}

\author{Dong Young Yoon, Soonchil Lee}

\affiliation{Department of Physics, Korea Advanced Institute of Science and Technology, Daejeon 305-701, Republic of Korea}

\author{Yoon Seok Oh, Kee Hoon Kim}

\affiliation{CeNSCMR, Department of Physics and Astronomy, Seoul National University, Seoul 151-747, South Korea}

\begin{abstract}

The canting angles and fluctuation of the magnetic ion spins of
spinel oxide MnCr${}_2$O${}_4$ were studied by nuclear magnetic
resonance (NMR) at low temperatures, which has a collinear
ferrimagnetic order below $T_C$ and a ferrimagnetic spiral order
below $T_s < T_C$. Contrary to previous reports, only one spin
canting angle of Cr ions was observed. The spin canting angles of
Mn and Cr ions in the ferrimagnetic spiral obtained at a liquid-He
temperature were $43\,^{\circ}$ and $110\,^{\circ}$, respectively. The
nuclear spin-spin relaxation was determined by the Suhl-Nakamura
interaction at low temperatures but the relaxation rate $T_2^{-1}$ increases
rapidly as the temperature approaches $T_s$. This indicates that the fluctuation of the spiral
component becomes faster as the temperature increases but not fast enough to leave an averaged hyperfine field to nuclei in the time scale of nuclear
spin precession in the ferrimagnetic phase, which is on the order of
$10^{-8}$ s. The spiral volume fraction measured for various
temperatures reveals that the collinear and the spiral
ferrimagnetic phases are mixed below the transition temperature of
the spiral order. The temperature hysteresis in the
volume fraction implies that this transition has first-order characteristics.

\end{abstract}

\maketitle

\section{Introduction}

One of the reasons spinel oxides AB${}_{2}$O${}_{4}$ have been of
interest is the various spin structures they show at low
temperatures.~\cite{M. Schmidt, N. Tristan, V. O. Garlea} Cubic spinels
\emph{M}Cr${}_2$O${}_4$ (\emph{M}=Mn, Co) have drawn much
attention recently in relation with multiferroics since it was
found that their spin configuration leads ferroelectric.~\cite{Y. Yamasaki} The
magnetic phase of \emph{M}Cr${}_2$O${}_4$ changes from
paramagnetic to ferrimagnetic and from ferrimagnetic to spiral
orders as the temperature decreases, even becoming glass-like at low
temperatures. The complexity in the spin structure of cubic
spinels results from geometrical frustration among the spins
of B site ions, which form the lattice of corner-sharing tetrahedra.
\emph{M} (Cr) ions occupy the A (B) sites that locate at the center of a
tetrahedron (octahedron) in the oxygen lattice. Both the \emph{M} ions
and Cr ions are magnetic, and their magnetization
directions are opposed in the ordered states.

The antiferromagnetic exchange interaction $J_{BB}$ among the
spins of the B site ions does not induce ordered states due to
the  aforementioned frustration.~\cite{P. W. ANDERSON} However, the addition of the
exchange interaction $J_{AB}$ between the A and B site ions can
give rise to a magnetic order. The relative strength of $J_{AB}$
to $J_{BB}$ is the key factor determining the state of the
magnetic order, including the spin canting angle. The N\'{e}el
configuration, a collinear ferrimagnet, is expected to be the
ground state when $J_{BB}$ is much weaker than $J_{AB}$. As
$J_{BB}$ increases, both the \emph{M} and Cr spins become canted and the
magnetic structure becomes complex. A classical theory predicts that the
ground state of a cubic spinel is determined by the parameter $u = 4
J_{BB}S_{B}/3 J_{AB}S_{A}$, where $S_A$ and
$S_B$ are the spin magnitudes at the A and B sites, respectively.~\cite{D. H. LYONS}
The N\'{e}el configuration is the ground state when $u<8/9$ and the
ferrimagnetic spiral is locally stable when  $8/9< u < 1.3$. It becomes unstable when $u>1.3$.

An early neutron diffraction experiment showed that the ferrimagnetic spiral is the ground spin state of
the cubic spinel MnCr${}_2$O${}_4$.~\cite{J. M. HASTINGS} Two different spin canting angle values were measured for the Cr ions, while only one value was
measured for the Mn ions. The cone angles of the Mn ions, Cr(I), and
Cr(II) in the ferrimagnetic spiral were $24\,^{\circ}$, $104
\,^{\circ}$, and $152\,^{\circ}$, respectively. The approximate
$u$ value estimated by these measurements is 1.6, which predicts
that the ferrimagnetic spiral is unstable. In contrast, the cone
angle of Mn ion spins measured by NMR was $63\,^{\circ}$
or $42\,^{\circ}$, while those of Cr(I)
and Cr(II) spins were $94\,^{\circ}$ and $97\,^{\circ}$, respectively.~\cite{T. W. HOUSTON,T. Tsuda,H. NAGASAWA}
Theory poorly matches these numbers,
but it predicts a very unstable spiral state in general. The question as to whether the ferrimagnetic spiral is the stable ground state in the
cubic spinels was raised again by a recent neutron diffraction
experiment involving $\mathrm{MnCr_2O_4}$.~\cite{K. Tomiyasu} It was
reported that the ferrimagnetic state is long-range ordered for
all temperatures below $T_C \sim 50$ K, and that the spiral component
appears in the plane orthogonal to the direction of the
ferrimagnetic order below $T_s \sim 20$ K. However, it is only short-range ordered.

The present study investigates the characteristics of the ordered
spin state of $\mathrm{MnCr_{2}O_{4}}$ at a low temperature by nuclear magnetic
resonance (NMR). First, the cone angles of Mn
and Cr ion spins, to which the previous NMR and neutron diffraction
results gave different values, were measured. Information  pertaining to accurate
cone angles is important to understand not only the ground state
of the spin order but also the electrical polarization.~\cite{Hosho Katsura} We measured
the nuclear spin-spin relaxation time $T_2$ as a function of the
temperature to study the change in the spin fluctuation with
temperature. The result indicates that the fluctuation of the
spiral component increases rapidly as the temperature
approaches the phase transition temperature of the spiral order
$T_{s}$, above which the fluctuation is too fast for the spiral
component to be measured by even local probes such as neutron
diffraction or NMR. The temperature-dependence of the NMR signal
intensity reveals that the spiral phase is mixed with the
ferrimagnetic phase at temperatures below $T_{s}$.

\section{EXPERIMENT}

A polycrystalline MnCr${}_2$O${}_4$ sample  was synthesized by a solid-state reaction from a molar ratio mixture of MnO and
Cr${}_2$O${}_3$ powders. The mixture was sintered at
$1100\,^{\circ}$C for 12 hours in an Ar environment, for 12 additional hours at $1200\,^{\circ}$C, and finally for 24 hours at
$1300\,^{\circ}$C. A single-crystalline sample was grown by the
flux method using a mixture of MnCO${}_3$, Cr${}_2$O${}_3$,
PbF${}_2$ and PbO (molar ratio = 1:1:2:2). NMR signals were obtained
by the conventional spin echo method using a custom-made spectrometer in the temperature range of 4 to 20 K. To estimate the spin canting angle, the resonance frequency was
measured for various magnetic fields up to 4 T. The nuclear
spin-spin relaxation time, $T_2$, was obtained by varying the time delay
between the $90\, ^{\circ}$ and $180\, ^{\circ}$ pulses. The ${}^{53}$Cr NMR
spectrum was measured in the frequency range from 60 to 70 MHz,
and the ${}^{55}$Mn NMR spectrum was assessed from 530 to 560 MHz. As the
spectral width was very broad, the signal intensity was measured
as a function of the frequency after selective excitation.

\section{RESULTS AND DISCUSSION}

\begin{figure}
\includegraphics[width=1\textwidth]{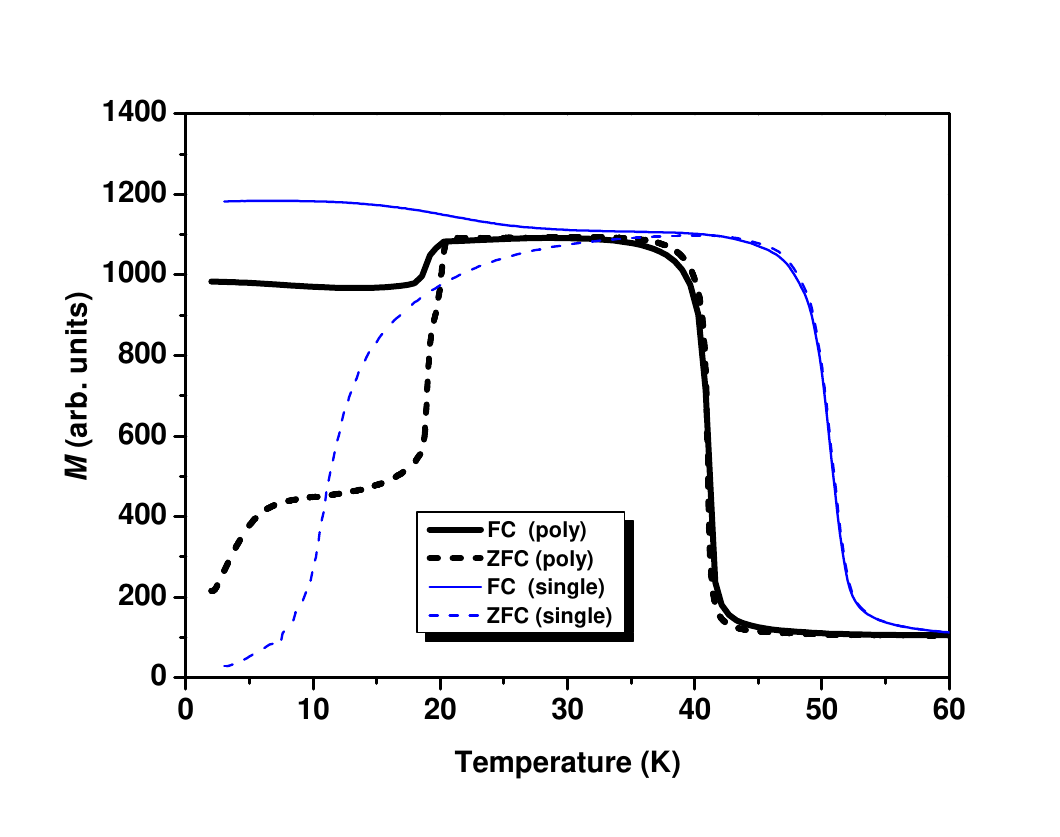}
\caption{(Color online) Magnetization vs. temperature curves obtained at 50 Oe: the thick solid and dashed lines (black) represent the FC and ZFC M(T) curves of the polycrystalline sample, respectively, and the thin lines (blue) represent those of the single crystal.}
\label{Magnetizationcurve}
\end{figure}

The magnetization versus the temperature curves show a discrepancy in
the transition temperatures of the ferrimagnetic spiral and the
collinear ferrimagnet of the polycrystalline and single-crystalline samples. In Fig.~\ref{Magnetizationcurve}, the thick solid and dashed lines represent the field cooling (FC) and zero
field cooling (ZFC) M(T) curves of the polycrystalline sample,
respectively, and the thin solid and dotted lines represent those
of the single crystal. $T_\mathrm{C}$ is relatively well defined
by the abrupt increase in the magnetization in both samples of approximately 40 K for the polycrystalline sample and 50 K for the single-crystalline sample. $T_{s}$ of the polycrystalline sample is more
clearly defined by the abrupt decrease in the magnetization at 20 K
compared to that of the single-crystalline sample, whose magnetization decreases smoothly at approximately 12 K only in the ZFC case. The M(T) curves
of our polycrystalline and single-crystalline samples are in
good agreement with those in previous reports.~\cite{K. Tomiyasu, Winkler} One
of the reasons for the difference in the characteristics of the two
samples is the site disorder in the single-crystalline sample.
X-ray absorption spectroscopy showed that the Mn ions occupy only
the A sites and the Cr ions occupy the B sites in our polycrystalline sample,
whereas both ions are found in both sites in the single-crystalline
sample.~\cite{J.H. Park} All of the experimental data described below
were obtained from the polycrystalline sample.

\begin{figure}
\includegraphics[width=1\textwidth]{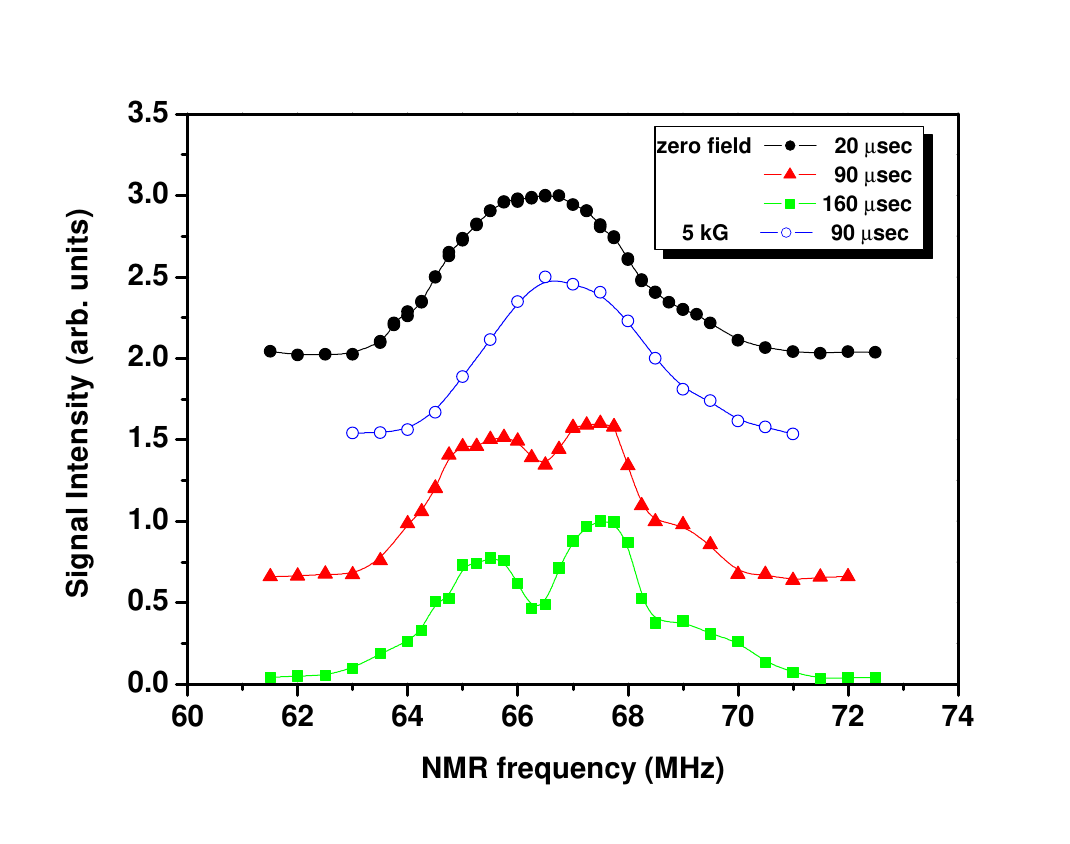}
\caption{(Color online) The filled circles represent the zero-field Cr NMR spectrum obtained at the echo time of 20, 90, and 160 $\mu s$, and the open circles represent the Cr NMR spectrum obtained in the external field of 5 kG and at the echo time of 90 $\mu s$.}
\label{zerofield}
\end{figure}

Figure~\ref{zerofield} shows the ${}^{53}$Cr NMR spectrum
obtained in a zero field at 6.5 K for several different echo times.
The spectrum obtained at the echo time of $20~\mu s$ shows a very
well-defined single peak whose width is about 5 MHz. The single
peak centered around 67 MHz at a short echo time of $20~\mu s$ appears to split into a double peak as the echo time supasses
$90~\mu s$. This is not a splitting but a suppression of the
spectral intensity around the center due to the frequency-dependent nuclear spin-spin relaxation rate. In an ordered
magnetic insulator containing a high concentration of identical
magnetic nuclear spins, the Suhl-Nakmura (SN) interaction, in
which nuclear spins are indirectly coupled by virtual magnons, is
expected to play a major role in the NMR relaxation at low
temperatures. It is known that the SN interaction generates a field-
and frequency-dependent $T_2$ with a minimum occurring in the
center of the spectrum, as the majority of nuclear spins
precess at this frequency.~\cite{J. Barak,R.R. ARONS} The double peak
feature of the spectrum obtained at $90~\mu s$ disappears in the
spectrum obtained at the same echo time, however, in a magnetic field,
as denoted by the open circles in the figure. This is consistent with the fact that the spin-spin relaxation rate due
to the SN interaction decreases as the field increases. The
dependence of $T_2$ on the frequency and field confirms that the SN
interaction is the main source of Cr nuclear interaction in
MnCr${}_2$O${}_4$. This most likely explains why a double peak
was observed in the previous NMR work, rather
than the difference in the sample quality.~\cite{H. NAGASAWA} The difference of only 3 $\%$ in
the canting angles of the two Cr spins associated with the two peaks
in the previous NMR report supports this claim, because the
experimental error of the canting angle as estimated by NMR is larger
than this in general. The ${}^{55}$Mn NMR spectrum showed a well-defined single peak centered around 550 MHz in a zero field at the liquid-Helium temperature.

The spin canting angles of the Mn$^{2+}$ and Cr$^{3+}$ ions
relative to the magnetization direction are determined by the
shift of the spectrum with an external field. The NMR resonance
frequency $f$ is proportional to the magnitude of the total field, which is the vector sum of the hyperfine field $H_{\mathrm{hf}}$
and external field $H_{\mathrm{ext}}$. This is expressed as follows:
\begin{eqnarray}
\ f &=& \gamma / 2\pi \left| \overrightarrow{H}_{\mathrm{ext}}+\overrightarrow{H}_{\mathrm{hf}} \right|\nonumber \\
  &\simeq& \gamma / 2\pi \left( H_{\mathrm{hf}} - H_{\mathrm{ext}}\cos\theta  \right),\nonumber
\end{eqnarray}
where $\gamma$ is the gyromagnetic ratio, and $\theta$ is the
angle between $H_{\mathrm{hf}}$ and $H_{\mathrm{ext}}$. The
direction of the hyperfine field is antiparallel to the local
magnetization in most magnetic materials, providing the minus sign in
the equation. As the hyperfine fields of the Mn$^{2+}$ and
Cr$^{3+}$ ions in MnCr${}_2$O${}_4$ are more than one order of
magnitude larger than the external magnetic field used in the
experiment, the first-order approximation of the total field can
be taken. The slope of the frequency shift with external
field is then determined as $\gamma /2\pi \cos\theta$.

\begin{figure}
\includegraphics[width=1\textwidth]{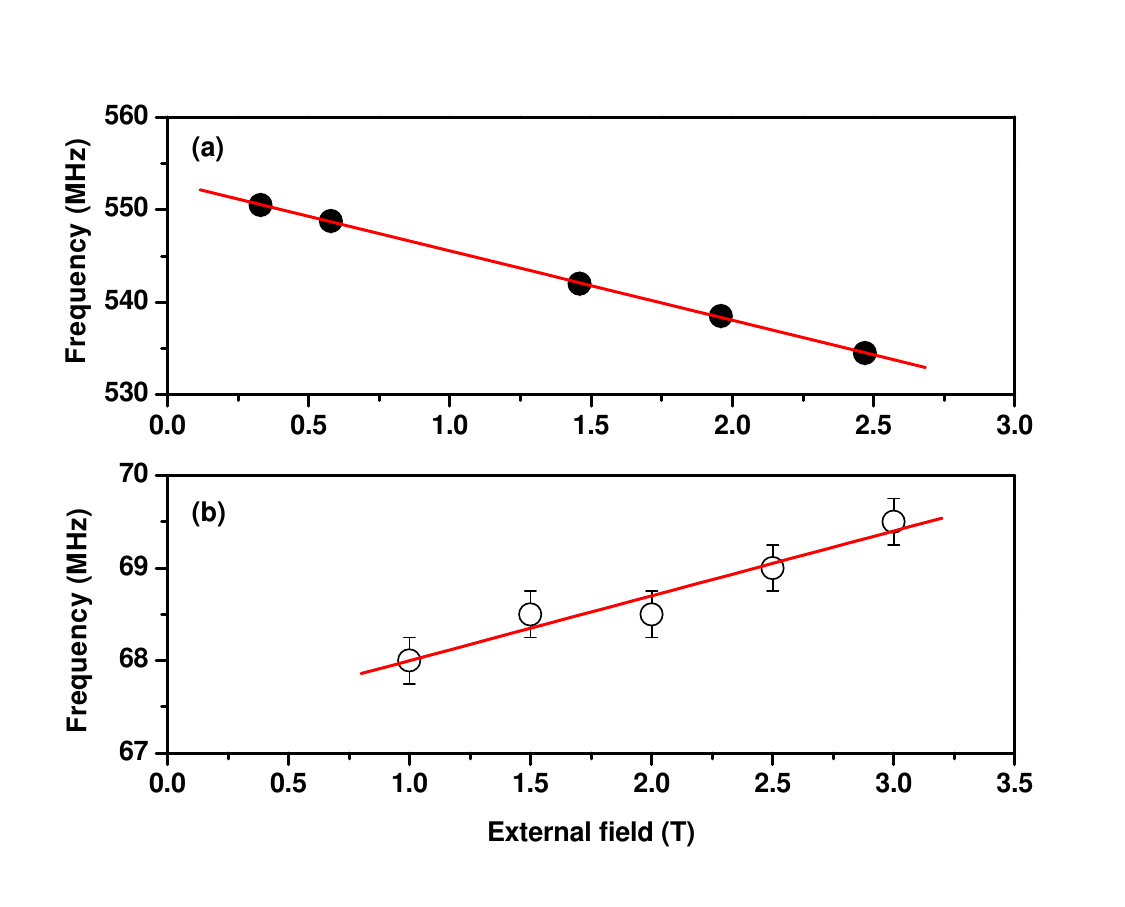}
\caption{(Color online) (a) The filled circles represent the central frequency of the Mn NMR spectrum obtained in the external field at 4.2 K. (b) The open circles represent the central frequency of the Cr NMR spectrum obtained in the external field at 4.2 K. The red lines denote the linear fit, of which the slope is $\gamma/2\pi \cos \theta$.}
\label{coneangle}
\end{figure}

In Fig.~\ref{coneangle}, the center frequencies of the Cr and Mn
NMR spectra obtained at 4.2 K are plotted as a function of the
external field. Both of the frequencies change linearly in the
experimental field range, as expected. The resonance frequency of
the Mn spectrum decreases as the field increases, whereas that of the
Cr spectrum increases. This visually shows that the spin directions
of the Mn and Cr ions are opposite to each other, because the
signs of the hyperfine fields of the ions are identical. From the
slope of the linear fit to the data, the spin canting angles of
the Mn and Cr ions were determined to be $43\pm5\,^{\circ}$ and
$110\pm5\,^{\circ}$, respectively. This is contrary to the
previous neutron diffraction or NMR experiments that reported two
different values of spin canting angles for Cr ions. The canting
angle of the Mn spin is identical to that in one of the previous NMR
reports~\cite{T. Tsuda} but twice as large as the value of the
neutron result~\cite{J. M. HASTINGS}. The canting angle of the Cr
spin is consistent with the previous NMR measurement~\cite{H. NAGASAWA} and one of the values given by the neutron diffraction~\cite{J. M. HASTINGS}.
Considering that no single value of $u$ can result in these cone angles, the
classical theory fails to explain the result. However, both of the
values corresponding to the Mn and Cr spin canting angles indicate
that the ferrimagnetic spiral configuration is unstable in
MnCr${}_2$O${}_4$. This reminds us of the fact that a long range
order of the ferrimagnetic component accompanies a short range
order of the spiral component below $T_s$ in MnCr${}_2$O${}_4$.\cite{K. Tomiyasu}

\begin{figure}
\includegraphics[width=1\textwidth]{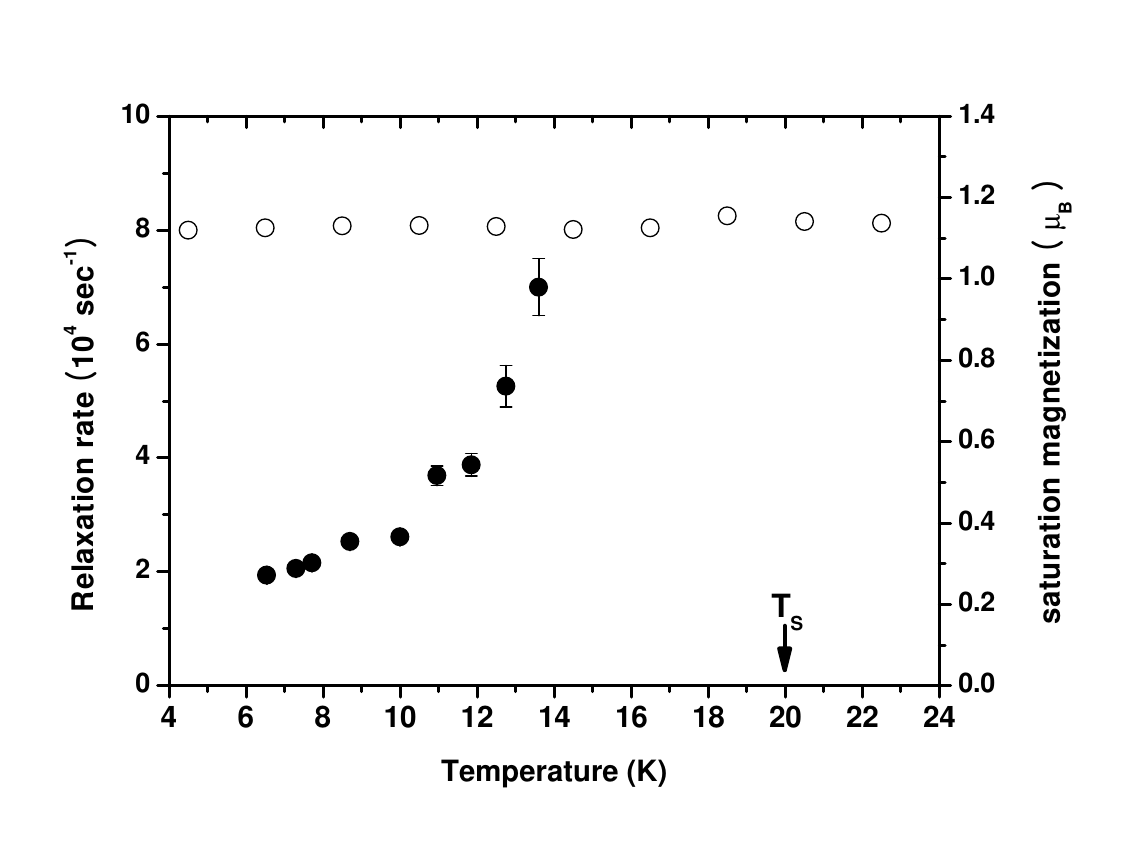}
\caption{The filled circles  show the relaxation rate $T_2^{-1}$ at the central frequency of the Cr NMR spectrum with the temperature. The open circles denote the saturation magnetization.}
\label{T2}
\end{figure}

Figure~\ref{T2} shows the nuclear spin-spin relaxation rate
$T_2^{-1}$ of Cr ions at temperatures ranging from 6.5 K to 14.5
K. The relaxation rate increases relatively slowly with the temperature below 11
K, above which the slope becomes steep. The main interaction a Cr ion nucleus experiences is the interaction with the electron spins of magnetic ions and the prominent relaxation source is SN interaction that is mediated by spin wave as mentioned above. The relaxation due to SN interaction is normally temperature independent. Therefore, the weakly temperature dependent relaxation with the rate of $2 \times  10^{4} ~\mathrm{sec}^{-1} $ near 7 K should be mainly due to SN interaction. The additional relaxation increasing with temperature indicates that the spin fluctuation becomes too large to be described in the framework of spin waves as the temperature approaches $T_s$. The previous observation that the spiral component of the spin order is unstable and short ordered implies that it is the spiral component of the Cr ion spins that fluctuates. This interpretation is also consistent with the saturation magnetization plotted together with the nuclear relaxation rate in Fig. \ref{T2}.  The value of the saturation magnetization stays at $1.1~{\mu}_{B}$ independent of temperature even when the temperature crosses $T_s$, where the spiral component is generated or vanishes. This means that the component along the easy axis remains the same while the spiral component perpendicular to it is averaged out by fluctuation crossing $T_s$, leaving only the ferrimagnetic order.

\begin{figure}
\includegraphics[width=1\textwidth]{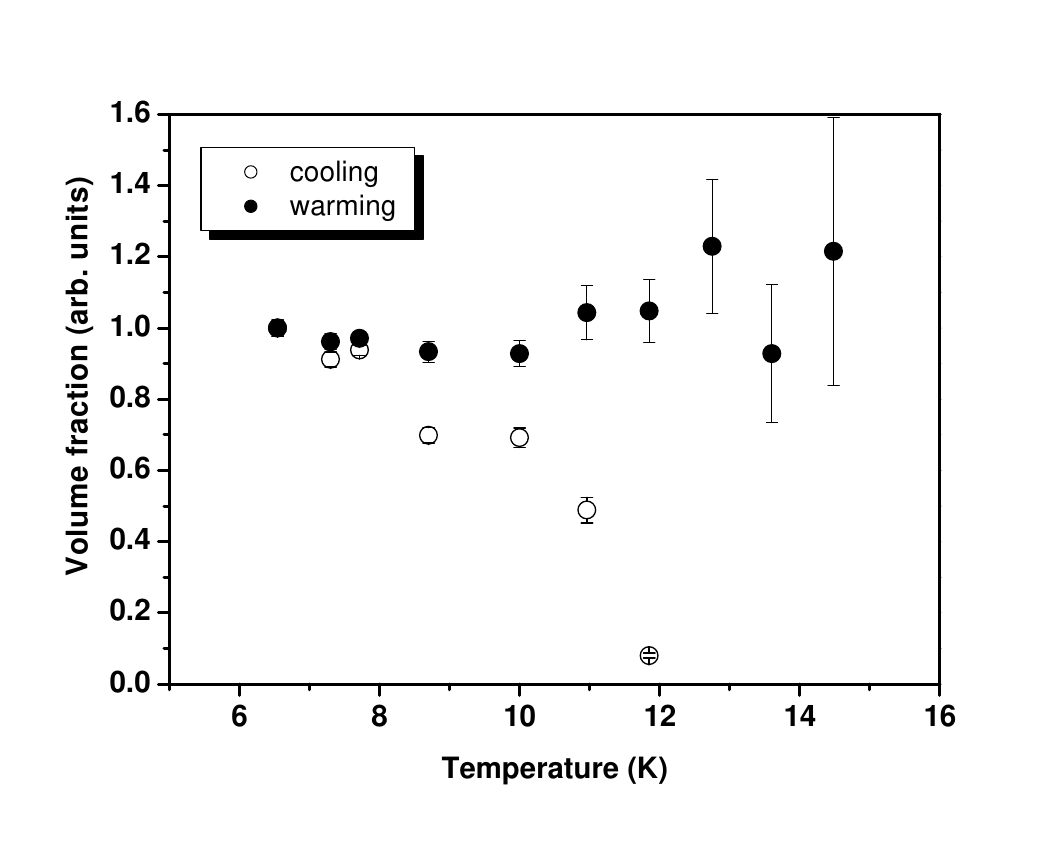}
\caption{The volume fraction of the ferrimagnetic spiral phase vs. the temperature. The open circles represent the volume fraction of the ferrimagnetic spiral obtained while cooling, and the filled circles represent that obtained while warming.}
\label{domain}
\end{figure}

The NMR signal intensity is in general a function of the temperature,
$T_2$, and the number of nuclei. The data points in
Fig.~\ref{domain}, where the Cr NMR signal intensity vs. the
temperature is plotted, were obtained from the raw experimental
data after temperature and $T_2$ correction. Thus, they represent
the number of the nuclei producing the signal. The signal
intensity obtained while warming the sample stays constant while
that obtained while cooling it changes. It is worthwhile to note that
the NMR signal is observed not in the ferrimagnetic phase but only in the spiral phase of MnCr${}_2$O${}_4$. Therefore,
the corrected signal intensity in the figure is proportional to
the volume of the ferrimagnetic spiral phase. Upon warming, the entire volume of the MnCr${}_2$O${}_4$ sample maintains its ferrimagnetic spiral phase until the change to the ferrimagnetic
phase at $T_s$. Upon cooling, however, MnCr${}_2$O${}_4$ remains in
the ferrimagnetic phase well below $T_s$. In the temperature
region where the signal intensity changes, the two phases coexist.
Depending on the temperature change history, the ferrimagnetic spiral
phase is embedded in the matrix of the collinear ferrimagnet
phase. This mixed phase might have caused some error in the measurement of
various quantities in the previous neutron diffraction and NMR
experiments. The temperature hysteresis in the volume of the
ferrimagnetic spiral phase can be ascribed to a first-order
transition at $T_s$. An ESR work on the similar cubic spinel
CoCr${}_{2}$O${}_{4}$ showed an abrupt shift of the frequency
at $T_{s}$, also indicating a first-order transition.~\cite{T. A.
KAPLAN} The experimental evidence is in conflict to the second-order transition on which the classical theory is based. The Mn NMR signal intensity also showed a similar temperature hysteresis.

\section{Conclusion}

The spin canting angles of Mn and Cr ions and the nuclear
spin-spin relaxation rates were measured in this study. Only one canting angle of Cr spins
was observed, contrary to that observed in previous neutron and NMR
experiments. The measured canting angles predict an unstable
ferrimagnetic spiral state at a low temperature. This instability is
consistent with the measurement of the
spin-spin relaxation rate. The relaxation rate increases more rapidly as the temperature increases until it appears to diverge at $T_s$. The rapid increase in the
relaxation rate near $T_s$ can be explained by the fluctuation of
the spiral component. The NMR signal is not observed due to this
strong relaxation near $T_s$. The NMR signal is unobservable
above $T_s$ as well, where the magnetic phase is collinear
ferrimagnetic. The magnetization remains the same, crossing $T_s$, indicating that the
canted spins in the ferrimagnetic spiral phase do not line up
along one direction, entering the ferrimagnetic phase; instead, the
fluctuation of the spiral component averages out to leave only the
magnetic component along the easy axis. The fluctuation accelerates as the temperature approaches $T_s$, and past $T_s$, it becomes fast enough
to make the spiral component unobservable in neutron
diffraction experiments but not fast enough to leave an averaged hyperfine field to
nuclei in the time scale of nuclear spin precession, which is on the
order of $10^{-8}$ s. The temperature hysteresis of the spiral
volume fraction indicates that the spiral and collinear
ferrimagnetic phases are generally mixed below $T_s$.

This work was supported by National Research Foundation of Korea (NRF) grants: (Nos. KRF-2008-313-c00290 and 2009-0078342).

\newpage

\end{document}